\documentclass{ws-procs11x85}
\usepackage{multicol,float} 

\usepackage{graphicx}
\usepackage{mathrsfs} 
\def\d{{\mathrm{d}}}

\begin{document}
\title{The quantum interest conjecture in (3+1)-dimensional Minkowski space.}

\author{Gabriel Abreu$^*$ and Matt Visser$^\dagger$}

\address{School of Mathematics, Statistics and Operation Research, Victoria University of Wellington, PO Box 600, \\Wellington, New Zealand\\
$^*$gabriel.abreu@msor.vuw.ac.nz\\$^\dagger$matt.visser@msor.vuw.ac.nz}

\begin{abstract}
The \emph{quantum inequalities}, and the closely related \emph{quantum interest conjecture}, impose restrictions on the distribution of the energy density measured by any time-like observer, potentially preventing the existence of exotic phenomena such as \emph{Alcubierre warp-drives} or \emph{traversable wormholes}. It has already been proved that both assertions can be reduced to statements concerning the existence or non-existence of bound states of a certain 1-dimensional quantum mechanical Hamiltonian. Using this approach, we will  informally review a simple variational proof of one version of the Quantum Interest conjecture in (3+1)-dimensional Minkowski space.  
\end{abstract}
\bodymatter
\begin{multicols}{2}
\section{Introduction}

Semiclassical general relativity predicts violations of the point-wise energy conditions \cite{Epstein&Roman} associated with negative energy phenomena such as warp drives \cite{Alcubierre&Pfenning}, traversable wormholes\cite{Worms}, and even \emph{time machines}\cite{TMachines}. However, the same theory can be used to constrain the magnitude and duration of negative energy pulses. Two of these restrictions are given by the quantum inequalities (QIs) and the quantum interest conjecture (QIC).

The QIs\cite{Ford:1990id} impose a lower bound on the expectation value (in a quantum state $\psi$) of the renormalized stress-energy tensor along a timelike geodesic, 
\begin{equation}
 I_{\psi,\omega}\equiv\oint\left<\,T_{00}^{ren}(t,0)\,\right>_{\psi}\omega(t)\,dt,
\end{equation}
weighted by a non-negative and normalized test function $\omega(t)$. The initial bounds on $I_{\psi,\omega}$ depended on the modified Bessel functions, and were obtained  using a Lorentzian test function\cite{Ford:1996er}. Eventually Flanagan found a more general bound in (1+1)-dimensional Minkowski space for a massless scalar field\cite{Flanagan:1997gn}, which does not depend on the specific choice of the test function,
\begin{equation}
\label{Flanagan}
 I_{\psi,f}\geq-\frac{1}{24\pi}\oint\frac{(\omega'(t))^2}{\omega(t)}\,dt.
\end{equation}
Similarly, Fewster and Eveson obtained bounds in (1+1) and (3+1) dimensional flat space\cite{Fewster:1998pu}. Although their two dimensional inequality is slightly weaker than (\ref{Flanagan}), their result for (3+1) dimensions is the most general and optimum bound for a massless scalar field,
\begin{equation}
\label{Few&Eve}
I_{\psi,f}\geq-\frac{1}{16\pi}\oint\left([\omega^{1/2}]''(t)\right)^2\,dt.
\end{equation}
Both inequalities, (\ref{Flanagan}) and (\ref{Few&Eve}), can be written as a more general statement for $2m$-dimensional spacetime\cite{Fewster:1999kr},
\begin{equation}
\label{QI}
 \oint \left<\,T_{00}^{ren}(t,0)\,\right>_{\psi}\,|f(t)|^2\,dt\geq-\frac{1}{c_m}\oint|D^{m}f(t)|^2\,dt,
\end{equation}
with $\omega(t)=f^2(t)$. Here $D$ is the derivative operator, and the set of constants $c_m$ are given by
\begin{eqnarray}
 c_m= \left\lbrace \begin{array}{cc}
6\pi &\quad m=1; \\
m\,\pi^{m-1/2}\,2^{2m}\,\Gamma(m-\frac{1}{2})&\quad m\geq 2.
\end{array} \right.
\end{eqnarray}
The test function is now normalized such that $\int_{-\infty}^{+\infty}|f(x)|^2\d x=1$. It is easy to check that we recover equations (\ref{Flanagan}) and (\ref{Few&Eve}) from (\ref{QI}), by setting $m=1$ and $m=2$, respectively. Furthermore, by integrating by parts, the QIs become a statement regarding the lack of negative eigenvalues for a one-dimensional pseudo-Hamiltonian,
\begin{equation}
\left<f|H|f\right>\geq0,
\end{equation}
where
\begin{equation}
\label{Ham}
H=(-1)^mD^{2m}+c_m\,V.
\end{equation}
Here $V\equiv\left<\,T_{00}^{ren}(t,0)\,\right>_{\psi}$ is effectively a \emph{potential} for a quantum mechanical system. Therefore it is possible to use the point of view of one-dimensional quantum mechanics (after some technical considerations\cite{Fewster:1999kr}) to reduce the QIs to a much \emph{simpler} framework.

\section{The QIC as an eigenvalue problem}

Once the quantum mechanical viewpoint is adopted, it is more convenient to also use its notation. Then the \emph{operator} $H$ can be written as $P^{2\,m}+ V$, where $P$ and $V$ are operators in the usual Hilbert space of square-integrable functions. This technical construction allows us to rewrite the eigenvalue problem in coordinates, as an ordinary differential equation for the eigenfunctions $\varphi$ (the test functions $f$ used before); the \emph{multiharmonic} time-independent Schr\"odinger equation (SDE), 
\begin{equation}
\label{multiSDE}
 (-1)^m\frac{\d^{2m}}{\d x^{2m}}\,\varphi(x)+V(x)\,\varphi(x)=E\,\varphi(x),
\end{equation}
where we have set $\hbar/2M\rightarrow1$ to simplify the algebra. Again, if $m=1$ we recover the (1+1)-dimensional case in the form of the time-independent SDE, 
\begin{equation}
\label{SDE}
 -\frac{\d^2}{\d x^2}\,\varphi(x)+V(x)\,\varphi(x)=E\,\varphi(x).
\end{equation}
Fortunately in this case, there is a theorem by Simon\cite{Simon:1976un}, which guarantees the existence of a negative eigenvalue for the SDE, if 
\begin{equation}
\label{Simon}
\int_{\infty}^{\infty}V(x)\,\d x\leq0.
\end{equation}

With this theorem, the QIC can be related to the QIs in (1+1) dimensions. It is also possible to reformulate the QIC for a more general set of energy pulses\cite{Teo:2002ne}, unlike the original formulation, which is restricted to $\delta$-function pulses\cite{Ford:1999qv}. To see this clearly, let us split the potential as
\begin{equation}
V(x)=V(x)_{+} - V(x)_{-},
\end{equation}
its positive part minus its negative part. Then in order to guarantee positive eigenvalues, the potential must violate (\ref{Simon}). That is,
\begin{equation}
\oint V(x)_{+}\,\d x>\oint V(x)_{-}\,\d x.
\end{equation}
Since here the potential $V$ represents the expectation value of the renormalized stress-energy tensor, it is clear that to fulfill the QIs the net energy density must always be positive, i.e., its positive part must always overcompensate the negative one. This is a simplified version of the  original formulation of the QIC\cite{Ford:1999qv}.
\section{The QIC in (3+1)-dimensional Minkowski space}

The (3+1)-dimensional case can be recovered from the multiharmonic SDE by choosing $m=2$. This yields the \emph{biharmonic} SDE, 
\begin{equation}
\label{bSDE}
\frac{\d^4}{\d x^4}\,\varphi(x)+V(x)\,\varphi(x)=E\,\varphi(x).
\end{equation}
However, we now need to generalize Simon's theorem for (\ref{bSDE}). 

First, via a variational argument, the lowest eigenvalue $E$ of (\ref{bSDE}) satisfies
\begin{equation}
\label{EigenVal}
E\leq\oint\left[\,\varphi''(x)^2+V(x)\;|\varphi(x)|^2\,\right] \d x,
\end{equation}
assuming all the test functions are normalized.

Secondly, let us choose the following test function 
\begin{equation}
\label{test}
\varphi_{test}=\sqrt{\frac{g\left(|x-\mu|/\sigma\right)}{\sigma}}.
\end{equation}
We then enforce the normalization
\begin{eqnarray}
 \oint \!g(|x|)\,\d x=1,\;\oint\! x\,g(|x|)\,\d x=0,\;\oint\! x^2\,g(|x|)\,\d x=1,\nonumber\\
 \!\!\!
\end{eqnarray}
to obtain
\begin{eqnarray}
\oint|\varphi(x)|^2\,\d x=1,\quad\oint x\;|\varphi(x)|^2\,\d x=\mu,
\end{eqnarray}
and
\begin{eqnarray}
\oint(x-\mu)^2\;|\varphi(x)|^2\;\d x=\sigma^2.\nonumber
\end{eqnarray}
Note that the kinetic term of (\ref{EigenVal}),
\begin{equation}
\label{Kinetic}
\oint\varphi''(x)^2\d x=\frac{\oint([\sqrt{g}(|x|)]'')^2\d x}{\sigma^4},
\end{equation}
diverges. It contains a  term proportional to $\delta$-function square, which arises from differentiating twice the absolute value. Nevertheless, by expanding the normalized functions $g(|x|)$ into a power series,
\begin{equation}
\label{gexpan}
g(|x|)=\sum_{n=0}^{\infty} a_n|x|^n,
\end{equation}
we can get rid of the troublesome term by properly setting the expansion coefficient $a_1$ to zero. Moreover, to make (\ref{Kinetic}) converge at zero, we also need $a_0>0$. The rest of the coefficients can be freely chosen. Then we have, from (\ref{EigenVal}), (\ref{test}) and (\ref{gexpan})
\begin{eqnarray}
\label{Expand}
 E\sigma^4&\leq&\!\kappa+a_0\,\sigma^3\oint V(x)\,\d x+a_2\,\sigma\oint |x-\mu|^2\,V(x)\,\d x\nonumber\\
&+&\!a_3\,\sigma\oint|x-\mu|^3\;V(x)\,\d x+O(1/\sigma),
\end{eqnarray}
with $\kappa=\oint([\sqrt{g}(|x|)]'')^2\d x$. Now choosing a sufficiently large $\sigma$, it is clear that $\oint V(x)\,\d x<0$ implies a negative eigenvalue for the differential equation.

Lastly, it is possible to collect more information from (\ref{Expand}) if we set $\oint V(x)\,\d x=0$. Then the next two terms become relevant. And since the sign and magnitude of $a_2$ and $a_3$ are arbitrary, either
\begin{equation}
 \forall\,\mu:\oint|x-\mu|^2\,V(x)\,\d x=0,
\end{equation}
or
\begin{equation}
\forall\,\mu:\oint |x-\mu|^3\,V(x)\,\d x=0,
\end{equation}
is a sufficient condition to guarantee the absence of a bound state. Differentiating twice the last expression with respect to $\mu$, we have
\begin{equation}
\forall \mu :\int_\mu^\infty V(x)\,\d x=0,
\end{equation}
and finally $V(x)=0$. That is, if $\oint V(x)\,\d x=0$, a necessary condition for the lack of a bound state is that $V(x)=0$. This proves the extension of Simon's theorem for the biharmonic SDE, and it also proves the QIC in (3+1)-dimensional Minkowski space.

To clarify the proof of the QIC, we must recover the notation of semiclassical general relativity. Then the version of the QIC we just proved states that the QIs imply, either $\left<\,T_{00}^{ren}(t,0)\,\right>_{\psi}\equiv0$ everywhere along the world line, \emph{or}
\begin{equation}
\label{AWEClike}
\oint \left<\,T_{00}^{ren}(t,0)\,\right>_{\psi}\,\d t >0,
\end{equation}
which is slightly stronger than the AWEC. Once again, splitting the energy density into its positive part minus its negative part, 
\begin{equation}
\oint \left<\,T_{00}^{ren}(t,0)\,\right>_{+}\,\d t >\oint \left<\,T_{00}^{ren}(t,0)\,\right>_{-}\,\d t,
\end{equation}
the positive energy density must overcompensate the negative part elsewhere along the world line.
\section{Discussion}
By proving the variant of Simon's theorem for the biharmonic SDE, we were able to reformulate the QIC in (3+1)-dimensional flat spacetime similarly to the (1+1)-dimensional case studied in references \refcite{Fewster:1999kr} and \refcite{Teo:2002ne}. In flat spacetime, an energy pulse which satisfies the QIs (and the QIC proved above) must also fulfill an AWEC-like inequality.

Although several technical aspects have not been mentioned here, a more detailed proof of the QIC in flat spacetime can be found in reference \refcite{Abreu:2008dh}.

\section*{Acknowledgements} 
This research was supported by the Marsden Fund administered by the Royal Society of New Zealand. \\
GA was additionally supported by Victoria University of Wellington. 
\end{multicols}

\end{document}